\documentclass[runningheads]{llncs}
\usepackage{graphicx}
\usepackage{amsmath}
\usepackage{fancyhdr}
\usepackage{makecell}

\fancypagestyle{firstpage}{
    \fancyhead[L]{}    
    \fancyhead[C]{This paper was presented as a poster at the 7th Annual Conference on Advances in Cognitive Systems (ACS), Cambridge, Aug 2019.}}

\begin{document}

%
\title{Cognitive Assistance for Inquiry-Based Modeling}

\author{Sungeun An\inst{1} \and
Robert Bates \inst{1} \and
Spencer Rugaber\inst{1}  \and
Jennifer Hammock\inst{2} \and
Emily Weigel \inst{3} \and
Ashok K. Goel\inst{1}
}
\authorrunning{S. An et al.}
%
\institute{School of Interactive Computing, Georgia Institute of Technology, Atlanta GA 30308, USA \and National Museum of Natural History, Smithsonian Institution, Washington, D.C. 20002, USA \and School of Biological Sciences, Georgia Institute of Technology, Atlanta, GA 30332 USA
\\
\email{sungeun.an@gatech.edu}}

\titlerunning{Cognitive Assistance for Inquiry-Based Modeling}
%
%
%

\maketitle              
\begin{abstract}
Inquiry-based modeling is essential to scientific practice. However, modeling is difficult for novice scientists in part due to limited domain-specific knowledge and quantitative skills. VERA is an interactive tool that helps users construct conceptual models of ecological phenomena, run them as simulations, and examine their predictions. VERA provides cognitive scaffolding for modeling by supplying access to large-scale domain knowledge. The VERA system was tested by college-level students in two different settings: a general ecology lecture course (N=91) at a large southeastern R1 university and a controlled experiment in a research laboratory (N=15). Both studies indicated that engaging students in ecological modeling through VERA helped them better understand basic biological concepts. The latter study additionally revealed that providing access to domain knowledge helped students build more complex models.

\end{abstract}
\thispagestyle{firstpage}
\section{Introduction}
In cognitive science, scientific discovery is sometimes considered as a general problem-solving process \cite{simon1992scientific}. To solve a problem, an agent must represent the problem, generate possible solutions to the problem, test them, and repeat the process if needed \cite{bridewell2006interactive}\cite{simon1992scientific}. In the course of scientific inquiry, the inquiry-based modeling approach is commonly used in which a scientist begins with a problem, generates multiple hypotheses, elaborates them into models, tests them through comparison with current observations, and revise them to better fit the observation \cite{clement2008creative}\cite{nersessian2010creating}. 

While playing an important role in scientific discovery and scientific inquiry \cite{bransford2000people}\cite{white1998inquiry}, the inquiry-based modeling is a knowledge rich domain in which novice scientists such as citizen scientists or students need much support. Novice scientists might not have a strong background in ecology or biology yet have an interest in learning and investigating a scientific question. For example, citizen scientists participate in scientific research in part to contribute to and expand the impacts of any study \cite{hand2010people}. Student scientists are concerned with dealing with authentic experiences in the real world, engaging in the dialogue used in the academic discipline, and thinking like a scientist \cite{shaffer1999thick}. 

Scientific modeling, however, requires domain knowledge (e.g., relationships between variables describing a system) and analytical skills (e.g. quantitative analysis). Without strong domain expertise, they tend to resort to “modeling fitting” bas the perceived task instead of the desired learning objectives, which is to gain a better understanding of the system as a while and the interactions between components in the system being modelled \cite{hogan2001cognitive}\cite{sins2005difficult}. Hence, there is a need for providing cognitive assistance for novice scientists to offload these cognitive tasks when engaged in scientific modelling. 

In this paper, we present a Virtual Ecological Research Assistant (VERA) that provides cognitive assistance for ecological modeling in two ways. First, it provides a visual language to represent conceptual models and a mechanism to evaluate the models through agent-based simulations without requiring any programming. Second, it provides access to biological knowledge through Smithsonian Institution’s Encyclopedia of Life (EOL; https://eol.org) \cite{parr2014encyclopedia} to help the novice scientist construct the conceptual models and set simulation parameters. Preliminary results from two experiments indicate that inquiry-based modeling using VERA helps college-level biology students better understand the biological principles underlying ecological phenomena. Further, access to structured ecological data in EOL helped the participants build more complex models. 

This paper consists of four sections. In the next session, we describe how VERA provides cognitive scaffolding for novice users. Next, we describe two studies: an experiment conducted in a real biology classroom and a more detailed experiment in our laboratory. In related work, we describe the theoretical frameworks that informed the design of our system. Lastly, we reflect on the results and discuss future directions for the research.

\section{VERA: Interactive Technology for Ecological Modeling}
\subsection{Virtual Laboratory for Experimentation}
\begin{figure}[ht!]
\centering
    \includegraphics[width=\textwidth]{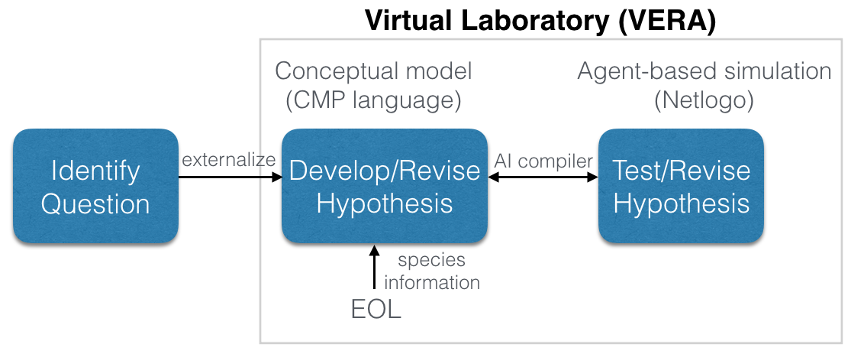}
    \caption{The scientific inquiry in VERA.}
    \label{scientific_inquiry}
\end{figure}

VERA acts as a virtual laboratory for inquiry-based scientific experimentation. Figure \ref{scientific_inquiry} illustrates the typical workflow using VERA. The user begins with a question. She then generates (potentially) multiple hypotheses for answering the question. In the process, the user may consult EOL for inspiration. Next, she elaborates on the hypotheses by constructing a detailed conceptual model. VERA provides an ontology of relationships derived from EOL. Then the user asks VERA to spawn a simulation from the conceptual model. VERA provides the user with the initial values for the parameters from EOL. VERA now automatically spawns the simulation and displays the results as graphs, for example, a graph indicating the changes in populations of various species over time. The user may now experiment with different simulation parameters, revise the conceptual model, or generate an alternative hypothesis.

Building on our previous efforts including the MILA-S system \cite{joyner2014mila}\cite{agarwal2018middle}\cite{an2018vera}, VERA is designed from the ground up to be a web-based application to provide broader access to the system, community building, and scalability (vera.cc.gatech.edu) whereas its predecessors were strictly desktop-based applications. The ontology of biotic and abiotic components and relationships in the conceptual models in the web version of VERA derives from Encyclopedia of Life (EOL; www.eol.org). VERA supports inquiry-based modeling by providing the user the authentic experience of a scientific inquiry (e.g., identifying a problem, proposing multiple hypotheses, testing the hypotheses, and rejecting/accepting the hypotheses) through construction, evaluation and revision of conceptual models. Testing a hypothesis is particularly important because then the user can take a more active role in constructing her own understanding in a feedback loop.

\subsection{Visual Language for Conceptual Models}
Conceptual models of ecological phenomena in VERA are expressed in VERA in the Component-Mechanism-Phenomenon (CMP) language \cite{joyner2014mila}. A CMP model consists of components and relationships between components. Borrowing from MILA-S and the CMP, component types for VERA are currently restricted to the following:
\begin{itemize}
    \item Biotic: A group of organisms to be modeled as a population with common parameters and behaviors.
    \item Abiotic: A non-biotic substance in the ecosystem being modeled that has an effect on other Biotic or Abiotic components.
    \item Habitat: A region in the model that represents a distinct space where Biotic and/or Abiotic components can exist and interact and even migrate between.
\end{itemize}

Possible interactions between biotic populations needed to be based on the ontologies used by EOL and its partner, Global Biotic Interactions (GloBI) \cite{poelen2014global}.  An interaction relates one component to another in a directed manner (e.g., component X consumes component Y), and consists of the following biotic interactions:  
\begin{itemize}
    \item X destroys Y: When X interacts with Y, it partially or wholly destroys a simulation entity of type Y with no carbon transfer to X.
    \item X produces Y: During the simulation, X will produce Y with some stochastic timing and amount.
    \item X consumes Y: When X interacts with Y, it will partially or wholly consume Y, with carbon transfer to X from Y.
    \item X becomes Y on death: When X expires, it produces Y.
    \item X affects Y: This is a generic interaction that provides growth rates (negative or positive) to modify Y when X interacts with it, where none of the above relationships apply.
\end{itemize}

\subsection{Integration with Encyclopedia of Life}
Integration with Encyclopedia of Life (EOL) provides access to vast amounts of structured and unstructured ecological data. VERA enables the user to look up species of interest and populate the simulation parameters available in EOL such as lifespan, body mass, offspring count, reproductive maturity, etc., which can be critical for building the most basic of models in a realistic fashion.

With component types and interactions defined, we consulted with the EOL team to identify what species attribute records in their database we should leverage to compile a conceptual model into a reasonable agent-based simulation.  Ultimately the joint team settled on the following critical attribute records for the first iteration of VERA:
\begin{itemize}
    \item Lifespan:  Average lifespan of individual organisms in this population, in months.
    \item Body mass:  Average body mass of individual organisms in this population, in kg.
    \item Carbon Biomass:  Average carbon biomass of individual organisms in this population, in kg.
    \item Respiratory rate: Rate of carbon biomass loss due to biological functions for this species, in kg/s.
    \item Photosynthesis rate: Carbon fixation via photosynthesis for this species, in kg/s.
    \item Assimilation efficiency: Percentage assimilation of consumed carbon biomass for this species, in percentage range [0.0, 1.0].
    \item Productive maturity: Average age to reproductive maturity for this species, in months.
    \item Reproductive interval: Average interval between reproduction cycles for this species, in months.
    \item Offspring count: Average number of offspring for this species during a reproduction event.
\end{itemize}

\subsection{AI Compiler for Generation of Agent-Based Simulations}
\begin{figure}[ht!]
\centering
    \includegraphics[width=\textwidth]{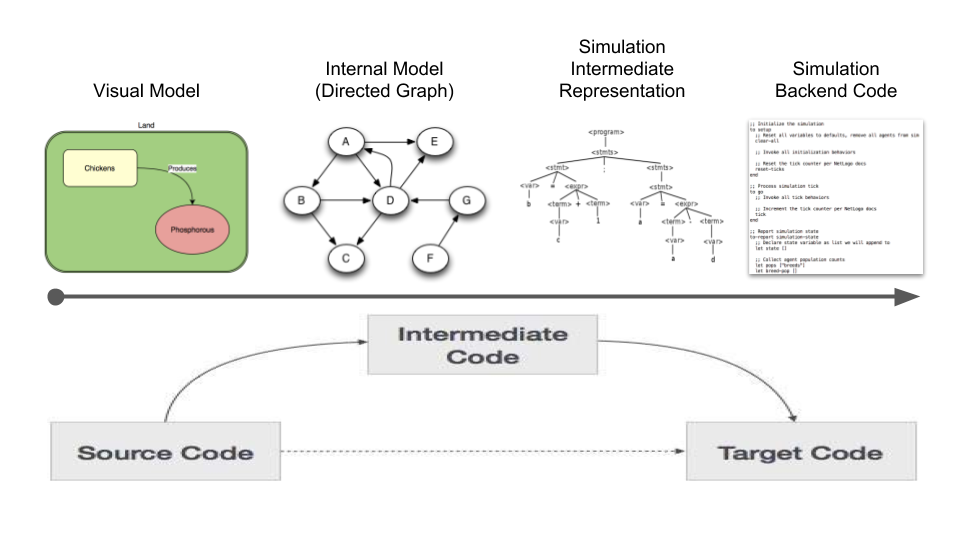}
    \caption{Mechanism of the visual conceptual model to a simulation.}
    \label{compiler}
\end{figure}

VERA currently leverages the same backend simulation system, NetLogo \cite{wilensky1999thinking} introduced in MILA-S \cite{joyner2014mila}), executing in a “headless” mode for server deployment. VERA uses an artificial intelligence compiler to automatically translate the conceptual models into the primitives of an agent-based simulation of NetLogo \cite{wilensky1999thinking}. Running the simulation enables the user to observe the evolution of the system variables over time and iterate through the generate-evaluate-revise loop. In this way, VERA integrates both qualitative reasoning in the conceptual model and quantitative reasoning in the simulation reasoning on one hand, and explanatory reasoning (via the conceptual model) and predictive reasoning (via the simulation) on the other. The mechanism by which the visual conceptual model is compiled into a simulation is demonstrated in Figure \ref{compiler}, borrowing from cross-compiler concepts.

\subsection{VERA and EOL in Action}
\begin{figure}[ht!]
\centering
    \includegraphics[width=\textwidth]{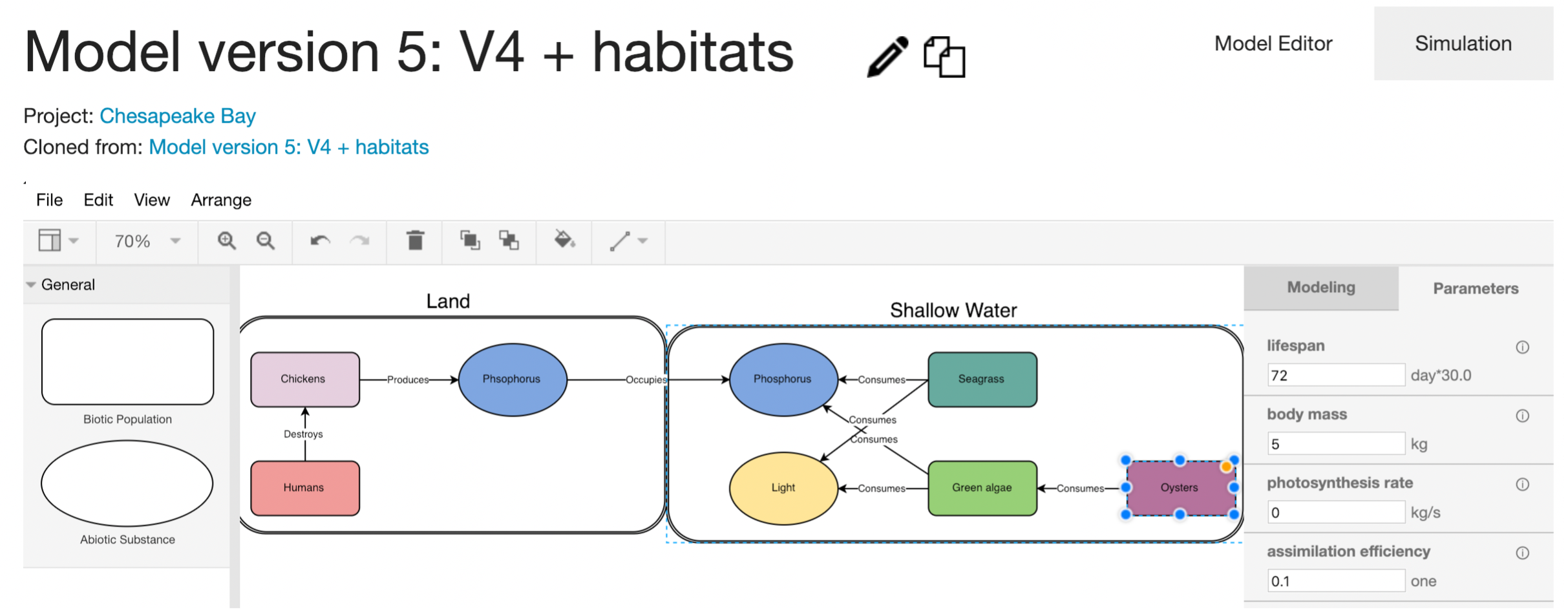}
    \caption{The screenshot of the VERA model editor page.}
    \label{model_editor}
\end{figure}

Figure \ref{model_editor} illustrates a CMP model of phosphorus run-off in the Chesapeake Bay; the large oval boxes in the middle depict habitats, in this case, land and shallow water. (The right side of the model editor depicts model parameters and their values.) The left panel contains model components for Biotic Populations and Abiotic Substances that can be added to the model.  The grid panel in the center is where the model is assembled.  The right panel contains a context-sensitive panel providing easy access to model, component, or interaction properties depending on what the user currently has selected.

In the following example, a citizen scientist is exploring the decline of pika (a small rodent-like creature) in the Rocky Mountains and wants to create a model of the observed food web. To leverage EOL to assist in model construction, they would first add a Biotic Population component by clicking once on the Biotic Population icon in the left panel, which prompts the user for a name for this population. Not being a domain expert, our user enters the common name “pika” and clicks on “Look up species on EOL” to locate the specific species.  The VERA system queries EOL for all matches to the scientific name or common name “pika” and checks for the existence of attribute records for each species, presenting all options.  Our user recognizes the common name “American pika” from discussions with fellow citizen scientists and selects the species which has 138 attribute records in EOL.

\begin{figure}[ht!]
\centering
    \includegraphics[width=\textwidth]{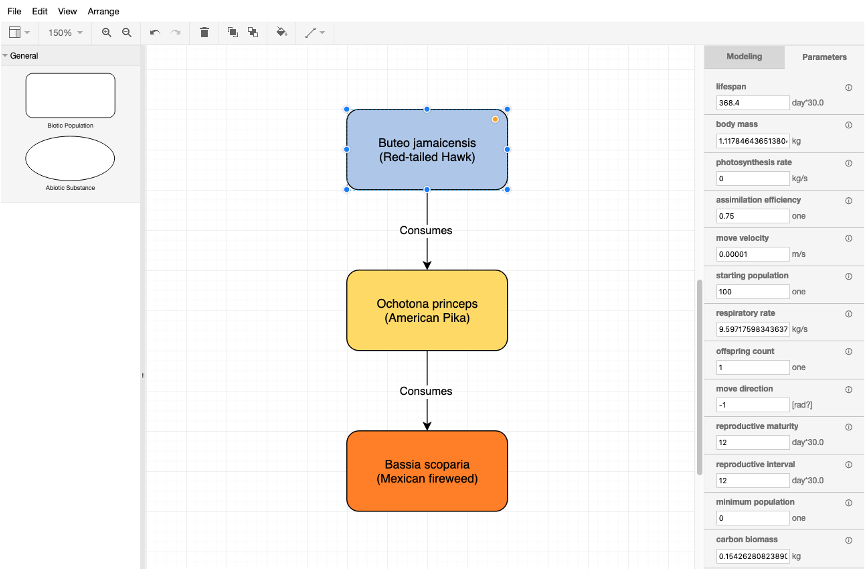}
    \caption{The food web of American pika.}
    \label{american_pika}
\end{figure}

In addition to retrieving the name from EOL, VERA also extracts the species attribute records from EOL that are relevant to the agent-based simulation as seen in the right panel in Figure \ref{american_pika}.  This provides a user with valuable data that a non-expert would be hard-pressed to locate and make sense of, reducing the cognitive load in model creation. Carrying on the model construction to add an observed predator (the Red-tailed hawk) and food source (Mexican fireweed) leveraging EOL lookups, along with adding consumption interactions between the populations, our intrepid citizen scientist has constructed a partial food web model revolving around the species of interest, as seen in Figure \ref{american_pika}. They can now experiment using simulation and model editing feedback loops within VERA to observe how changes in initial populations, food sources, and reproduction can affect the population counts in the ecosystem, to either support or refute their hypothesis regarding observed pika population changes.

\section{Study 1: General Ecology}
The VERA was introduced to a general ecology lecture course taught at a large southeastern R1 university in Fall 2018. This course is taught after the introductory biology course and uses a flipped format structure with regular in-class computer use. The study was conducted during two class periods after the students learned the basic information of “Energy flow and food webs.” The duration of each class period was 50 minutes. The purpose of this study was to see if engaging in VERA through exercise scenario can enhance students’ content knowledge in the ecological domain and understanding about the utilities of modeling and simulation. Also, since it was our first-time introducing VERA in a large population (N=91), we wanted to capture any possible usability issues and system errors that might happen during the study. 

\subsection{Demographics}
Among the registered students (N=91) in the class, 52 students completed the paired pre- and post-test, and 80 students submitted a reflection report. Among the 52 students, 22 were freshman; 18 were sophomore; 9 were junior; 1 was senior; 2 were others (e.g. GA). 11 students received credit for AP/IB biology from high school whereas another 41 students did not. The students’ self-assessments of their prior knowledge of biology and experience in modeling and interactive constructive games were modest; on a Likert scale, the average self-perceived familiarity with modeling was 2 (min=1; max=4); the average familiarity with games was 2.82 (min=1; max=5); and the average familiarity with biology was 4.92 (min=3, max=5).

\subsection{Study Design}
In the first-class period, the students were given 15 minutes to complete the pre-test which included nine questions pertaining to knowledge of ecology and modeling. Questions 1-5 covered ecological knowledge: they questioned a student’s ability to make predictions about food webs, graph visualization, inferring from data, inferring from graphs, and complex calculations. Questions 6-9 pertained to conceptual and simulation modeling. All questions were presented as multiple choice. The pretest was followed by a survey about student demographics and academic background. These questions included name, year, familiarity with biology, familiarity with modeling, and familiarity with constructive, world-building games (for example, SIMS, Minecraft). After the survey, the students were given a tutorial introduction to VERA for 20 minutes. The researcher walked through VERA step by step to explain how it works and what each simulation parameter meant.

In the second-class period, the students developed models of predator-prey interaction scenarios that involved wolf, sheep, and grass. After the exercise, the students were given another 15 minutes to complete a post-test identical to the pre-test. In the exercise, guided examples and packets were given to the student via Qualtrics to guide the exercise scenario. For example, the predator-prey interaction scenario was composed of four phases. The sequence of each phase formed a series of increasingly complex situations. The students started by trying to model a species with default propagation parameters and then investigate how changing these parameters, resource availability, and emergence of predators affect the species’ populations.

The first phase explored sheep and grass interaction where the sheep “consumes” grass, and grass was treated as an unlimited resource that never runs out. The second phase explored the interactions of sheep “consumes” grass and sunlight “affects” grass. While the grass was a limiting resource, the environmental factors, such as sunlight and rain, was introduced to control the grass population. In the third phase, the students explored a scenario where sheep “consumes” grass interacts with the controlled propagation parameters of the sheep. In the previous phases, the students used the default propagation parameters for the sheep population. In this phase, the students added real propagation parameters such as lifespan, reproductive maturity, and offspring count as in real ecology information retrieved from the EOL. The fourth phase was built on the phase 3 by adding a predator species, wolves, to the previous interaction. For each phase, the students followed the guided examples in the packet, and questions asked the students to write their simulation observations and why they thought these occurred. As the optional question, the students were asked to build their own model to control the growth of the sheep population possibly by adding more components or changing the parameter values. When building their models, they were encouraged to use species information in EOL’s TraitBank (eol.org/data\_search).

\subsection{Exercise}
\begin{figure}[ht!]
\centering
\begin{minipage}{0.35\textwidth}
        \centering
        \includegraphics[width=\textwidth]{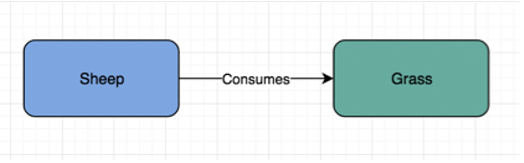}
    \end{minipage}\hfill
    \begin{minipage}{0.55\textwidth}
        \centering
        \includegraphics[width=\textwidth]{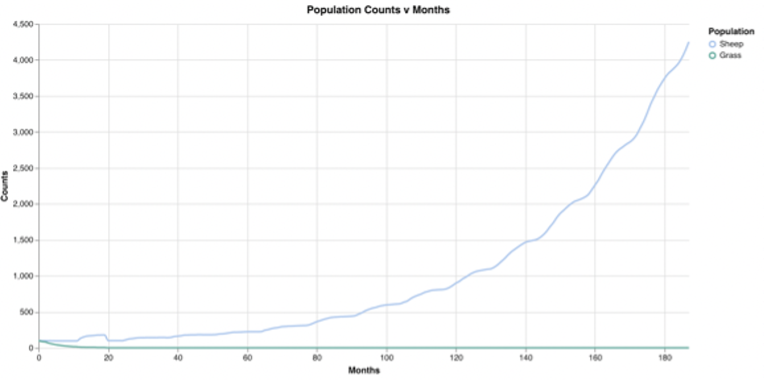}
        \\
    \end{minipage}
    \caption{The conceptual model and the simulation graph in the Phase 1.}
    \label{phase1_graph}
\end{figure}

The exercise scenario that was given to the students consisted of four phases. In phase 1, when the students simulated the model of the sheep alone, they saw that the sheep starved and died off. For example, the students observed that “This sheep population dies off quickly because it has no grass to eat” and “The sheep population completely dies off after 20 months because there is no food available.” When the grass population is added as an unlimited resource to be “consumed” by the sheep population, the students observed that the sheep population exploded (see Figure \ref{phase1_graph}). For example, students’ responses included “With grass being an unlimited resource, the sheep population continues grow exponentially” and “Sheep population decreases at first and then grows uncontrollably. Because grass is an unlimited resource, sheep have a continuous supply of food and can continue to grow as a population.”

In the phase 2, instead of making the grass population as unlimited resource, the students changed a parameter to make it a depleting resource like in a real ecology system. Consequently, the students saw that both the grass and sheep population decrease. For example, “Both populations die off as time goes on. This happens because the grass population is not unlimited unlike in phase 1.” To control the grass population, the sunlight was added to “affect” the growth rate of the grass positively. As a result, the students observed slower population growth compared to the previous phase. For example, the students’ responses were “Once sunlight was added, the grass population exponentially increased.”

In the phase 3, instead of using the default propagation parameters for the sheep, the students added real propagation parameters from EOL. Observing the population explosion in phases 1 and 2, the students saw a more controlled rate of the population growth of the sheep compared to the previous phases. The students wrote that “The grass population decreased a little over time, but the sheep population increased in a stair-step looking fashion” and “In this model with parameters, the sheep population begins low and steadily grows over time rather than starting with a greater population that quickly diminishes...”

\begin{figure}[ht!]
\centering
\begin{minipage}{0.45\textwidth}
        \centering
        \includegraphics[width=\textwidth]{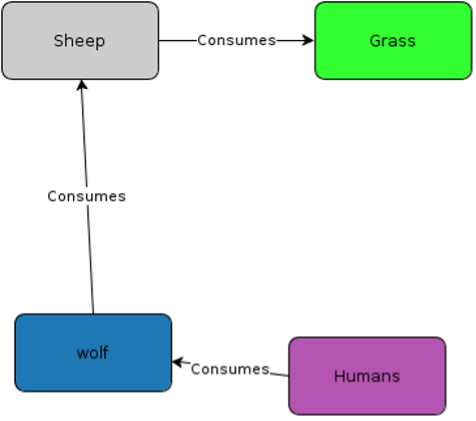}
    \end{minipage}\hfill
    \begin{minipage}{0.45\textwidth}
        \centering
        \includegraphics[width=\textwidth]{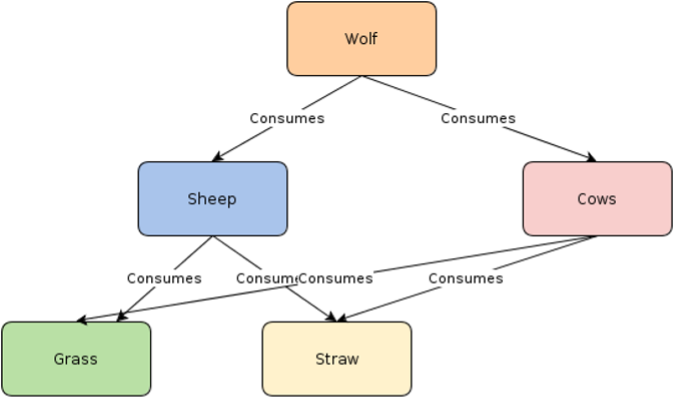}
        \\
    \end{minipage}
    \caption{The students’ models. Left: add predator. Right: add competitor and prey.}
    \label{student_models}
\end{figure}

In the final phase, the students added a predator on top of their models built in the previous phase. They put the real propagation parameters for the wolf from EOL as well. The students observed that the sheep population is influenced by the wolf population. They said, “By introducing the wolf population, the sheep population remains stable and only slightly decreases while the wolf population becomes quite large” and “With the presence of the wolf, predator to the sheep, the sheep population steadily decreases until it stabilizes at a very low count.” Some students created their own models to control the growth of sheep population. Their models added various types of components such as predators (human), prey/resources (straw), and competitors (goat, cows) (see Figure \ref{student_models}).

\subsection{Pre- and Post-Test}
The questions in the pre and post-test assessed the students’ abilities to reason about the biological relations and identify the utility of the conceptual and simulation model in general (see Table \ref{questions}). These abilities to make predictions and infer from data and graphs can be critical for building the most basic of models. 

\begin{table}[ht!]
\caption{Questions and types in the questionnaire}
\label{questions}
\begin{tabular}{p{0.05\linewidth}  p{0.3\linewidth}  p{0.6\linewidth}}  
    \hline
    \#  &  Type &  Question \\
    \hline
    1 & Making predictions about the food web & Which of the following would is most likely occur to populations within a community of organisms immediately after a wildfire burns their environment?  \\
    2 & Graph visualization & When resources are unlimited, which graph best represents the change in population size? \\
    3 & Graph visualization & When resources are limited, which graph best represents the change in population size? \\
    4 & Inferring from data & Which factors may stabilize a population explosion? \\
    5 & Inferring from graph & Which statement best describes what happened in 1850? \\
    6 & Complex calculation & Given the following information, what is the maximum number of offspring a species can produce during its lifetime? \\
    7 & Conceptual model & Based on your current knowledge, why might conceptual models be incorporated into the scientific method? \\
    8 & Simulation model & Based on your current knowledge, why might simulation models be incorporated into the scientific method? \\
    9 & Conceptual \& Simulation model & Why might a conceptual model paired with a simulation model of the same concepts be used as part of the scientific method? \\
    10 & Experiments via simulation & If you ran the same simulation multiple times, what would the simulator generate? \\
    \hline

    \hline\\
\end{tabular}

\end{table}

\subsection{Overall Effectiveness}
We used the responses of 52 participants on the pre- and post- tests. As shown in Table \ref{learning_gain}, the calculated t-value between the pre-test and the post-test on questions pertaining to biology is found to be significant (p value=0.0322 at 0.05 level). This shows the students significantly scored high achievement in the post-test when compared to the pre-test on the biology related questions. However, we did not observe similar gains on questions pertaining to conceptual and simulation modeling as parts of the scientific process (p value=0.501).

\begin{table}[ht!]
\caption{Overall results of the biological and modeling questions.}
\label{learning_gain}
\begin{tabular}{p{0.3\linewidth}  p{0.05\linewidth}  p{0.15\linewidth} p{0.15\linewidth} p{0.15\linewidth} p{0.15\linewidth}}  
    \hline
    Question & N & Estimate & Std. Error & t-value & Pr($>|t|$) \\
    \hline
    Biology & 52 & -0.4615 & 0.2125 & -2.171 & 0.0322* \\
    Model \& Simulation & 52 & -0.1154 & 0.1709 & -0.675 & 0.501
    \\

    \hline

    \hline\\
\end{tabular}
\end{table}

For Question 1, we observed an almost uniform correct response to the question in both the pre- and post-test, (N=51). Table \ref{learning_gain} shows that learning has occurred on Questions 2-5. Of the remaining questions which could show improvement, no one question appears to drive the overall effect of increase in biology-related knowledge. Questions 4 and 5 required understanding of how parameters and variables affect behaviors in the simulation model. This means that the students showed improvement in their understanding of how combinations of parameters such as lifespan and reproductive-rate contribute to the population size of a species. Typically, students do not get an opportunity to see how these parameters interact in class on biology or ecology; instead, these are taught as separate factors. 

On the relationship between students’ background knowledge and their learning gains, learning was independent of whether students had taken AP/IB biology, year, familiarity with modeling, and familiarity with biology. It was also independent of ASCI and MBVI data (math aptitudes). However, we found that familiarity in world- building games has a positive correlation with learning biological knowledge (Beta= 0.2367, SD=0.1007, t=2.35, p=0.0207*).

\subsection{Lessons Learned From the Study}
In the week following the study, we received students’ reflection reports about VERA. We received many positive statements about VERA (e.g. enjoyed, liked, cool, etc.). The student generally felt that the VERA system is interesting because it allows them to visually watch the interactions that involves complex variables. Most students felt that the parameters were difficult to deal with. For example, a student said, “The stumbling blocks for me were the parameters that we could set; I didn’t really understand how to manipulate them.”

Some students suggested providing pre-made models available that they can build on (“some possible improvements can be some pre-made models and/or relationships that you can then manipulate for your own,”); some suggested having a tutorial feature for explaining what parameters mean (“I feel like a short and simple description when hovering over each parameter would be a helpful improvement because I know there are so many elements of a population and its dynamics to investigate.”)

This study was conducted in the real classroom setting where we had less control on time spent on each section (e.g., some students came late in the classroom) and each student’s performance (e.g., some of the students’ computers did not work). This study was conducted with a large number of participants and provided evidence on the learning benefits of VERA as well as students’ positive comments. For better understanding about how students construct models, we wanted rich information about students’ models to see how the quality of models is related to students' demographics, learning gains, and how they use the system. Therefore, we conducted a second study in a more controlled manner with a relatively small number of participants.      

\section{Study 2: Lab Experiment}
Fifteen self-selected participants were recruited from a general biology introductory course taught at a large southeastern R1 university in Fall 2018. The experiment was conducted in two time slots on a single day in Fall 2018. The duration of the study was 2 hours for each participant. We reimbursed each participant in the study for their time with a gift card. Since the first study was conducted in a less controlled fashion with a large number of participations (N=91), the purpose of this study was to see 1) the effectiveness of VERA in enhancing ecological content knowledge and understanding about the modeling and simulation utilities, and 2) analyze students’ models in terms of complexity and creativity with a small number of participations (N=15) in more controlled fashion.

\subsection{Demographics}
Among the fifteen students, thirteen were freshmen whereas 1 student was a junior (1 student did not respond). One student indicated that he or she received credit for AP/IB biology from high school whereas the other fourteen students did not. The students’ self-assessments of their prior knowledge of biology and experience in modeling and interactive constructive games were modest. The average self-perceived familiarity with modeling was 2.26 (min=1; max=4); the average familiarity with games was 2.53 (min=1; max=4); and the average familiarity with biology was 2.66 (min=2, max=4).

\subsection{Study Design}
The study design followed the same structure as in the first study except for an open question session. Consequently, the study was composed of five procedures: pre-test, tutorial, exercise, open question, and post-test. We used the same questions for the pre- and post-test, the same tutorial and the same exercise as in the first study. Building on the exercise models, in the open question phase, students were given a scenario in which the population of sheep was decreasing and asked to explore the question: why does the sheep population decrease? This question addresses population changes, which is an important topic in ecology. It is akin to real-world ecological problems that professional scientists investigate, such as why are starfish dying off the west coast of USA, why is an algal bloom occurring in Chesapeake Bay, and why is the population of sea otters declining in Alaska? In our study, the students were asked to use VERA to devise multiple hypotheses that can explain the decline in the sheep population. They used VERA to build conceptual models from an explanatory hypothesis by adding new biotic or non-biotic components to the models and tweaking their parameter values. They then used VERA to generate agent-based simulations to evaluate their conceptual models and explanatory hypotheses. The students were also encouraged (but not required) to actively use the Encyclopedia of Life  biology knowledgebase when they need information about a given species. After the open-ended question, the students were asked to complete the post-test.

\subsection{Model Construction}
For the open-ended question, students came up with multiple hypotheses that can explain the decline in the sheep population. From the students’ answers, we identified six types of biotic and non-biotic components: 1) predator, 2) prey, 3) competitor, 4) virus/ parasite/ pesticide/ disease, 5) social factors such as hunting and demand for wool, and 6) environmental factors such as sunlight, wildfire, and pollution. Along with the six types of components, four types of relationships were used - consume, destroy, infect, and parasite - out of the eleven types of relationships that VERA supports. The most commonly used component was Predators (86.66\%; 13 out of 15 participants used it), followed by Prey, (60\%) Virus/Parasite, (40\%) Environmental factors (33.33\%), Social factors, (20\%) and Competitors (13.33\%). For the relationships, all participants used the Consume relationship in their model; Destroy was the second most commonly used relationship (46.66\%); Infect and Parasite of were each used by one participant (6.66\%).

When the participants were asked how frequently they used EOL when constructing their models, 3 participants responded they always used EOL; 6 participants answered most of the time; 4 participants answered about half the time; and 2 participants answered sometimes. No participants answered never. For the purposes of the usage (multiple answers were allowed for this question), the most common reason for using EOL was to find the parameter values such as lifespan, and birth-rate (12 participants), followed by finding the predators (10 participants), finding competitors (3 participants), and finding preys (2 participants).

\subsection{Pre- and Post-test}
For this study, we used the responses of fourteen participants on the pre- and post-tests (one participation did not respond to the post-test). As shown in Table \ref{learning_gain2}, the data indicates a significant increase in answering questions pertaining to biology (pre-test=71.61; post-test=83.33). We did not observe similar gains on questions pertaining to conceptual and simulation modeling as parts of the scientific process (pre-test=73.21; post-test=66.07).

\begin{table}[ht!]
\caption{Mean and standard deviation of pre-test and post-test results.}
\label{learning_gain2}
\begin{tabular}{p{0.25\linewidth}  p{0.2\linewidth}  p{0.2\linewidth} p{0.25\linewidth}}  
    \hline
    Question & N & Biology & Model \& Simulation \\
    \hline
    Pre-test & 14 & 71.61 (15.91) & 73.21 (13.52) \\
    Post-test & 14 & 83.33 (13.29) & 66.07 (19.66) \\
    \hline\\
\end{tabular}
\end{table}

\begin{figure}[ht!]
\centering
\includegraphics[width=0.95\textwidth]{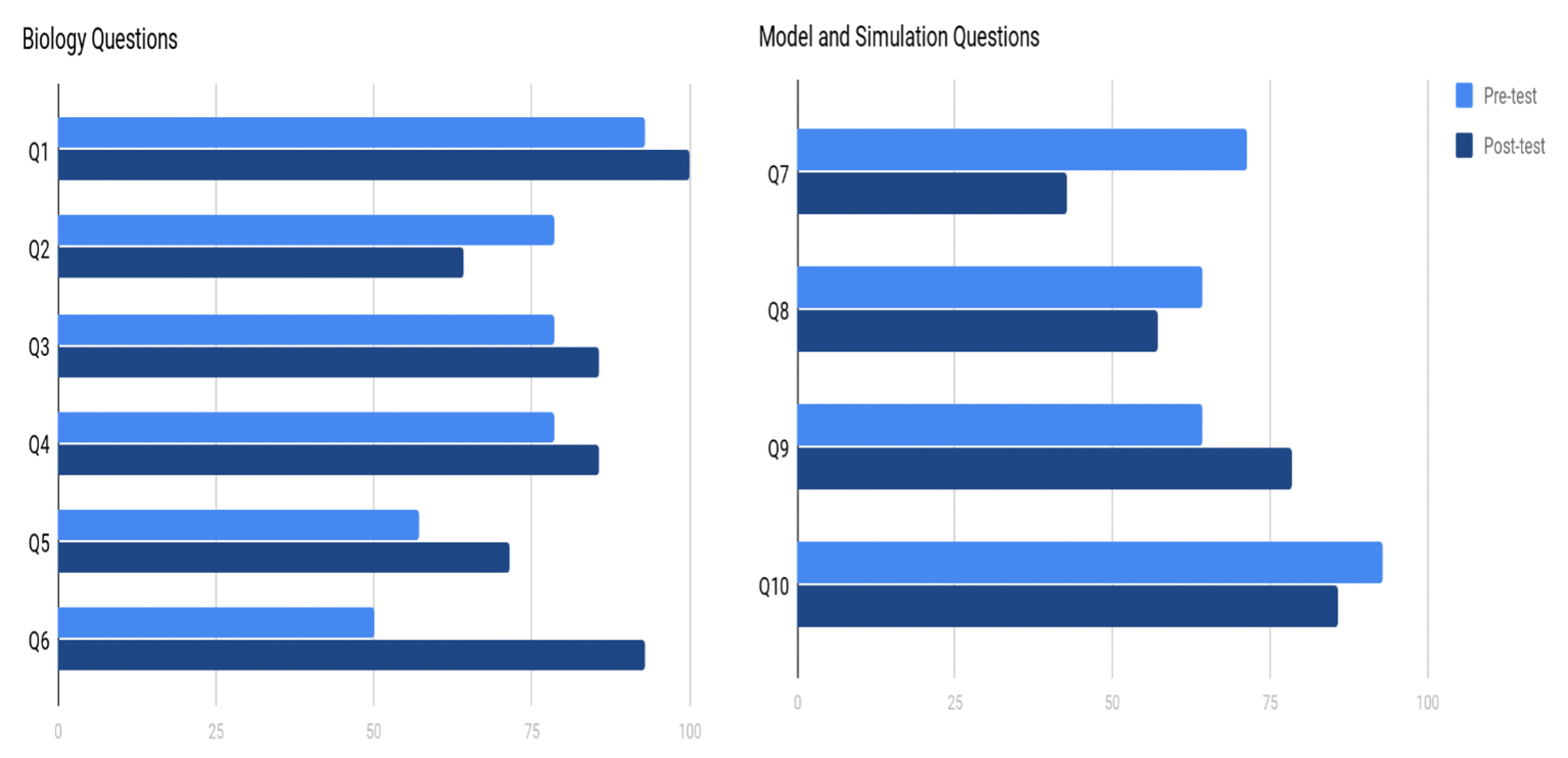}
\caption{The results of biology questions.}
\label{biology_questions}
\end{figure}

Figure \ref{biology_questions} (left) indicates that learning on the Questions 1, 3, and 4, and significant learning on Questions 5, and 6. Questions 5 and 6 required understanding of how parameters and variables affect behaviors in the simulation models (see Table \ref{questions}). This means that the students showed significant improvement in their understanding of how combinations of parameters such as lifespan and reproductive-rate contribute to the population size of a species. Typically, students do not get an opportunity to see how these parameters interact in class on biology or ecology; instead, they are taught as separate factors.

For the modeling and simulation questions (Questions 7-10), the learning occurred in Question 9 whereas learning did not occur for Question 7, 8, and 10 (see Figure \ref{biology_questions}-right). The learning in Question 9 indicates that the students were successful in understanding the integration of the conceptual model and the simulation as a scientific method as “it refines a hypothesis before conducting an empirical pilot study or experiment.” From their responses on Questions 7, 8, and 10, we posit that the students understand the conceptual and simulation models as “ground truths” rather than the abstractions of the world for explaining and making predictions about the world. For example, in Question 8, the most popular incorrect answer was that the simulation models can be incorporated into the scientific model “because they determine the exact outcome of an ecological scenario with set inputs.”

We found interesting correlations between students’ background knowledge and their learning gains. The learning gain was measured for each participant by subtracting each person’s pre-test score from their post-test score. We found that familiarity in modeling has a positive correlation with learning both ecological knowledge and the modeling process (r=0.228; r=0.441). Familiarity in interactive simulation games (e.g., Mine-craft) has a positive correlation with learning biological knowledge, but little correlation with learning about the modeling process (r=0.275; r=-0.050). Familiarity with biology has a negative correlation with learning biological knowledge (perhaps because they are already familiar with the biology content), and no correlation with learning about the modeling process (r=-0.170; r=0.044).

\subsection{Complexity and Creativity of Constructed Models }
The participants came up with 2.66 hypotheses on average (min=2, max=4). They used VERA to elaborate each hypothesis into a conceptual model, and then to test and revise the model. We analyzed the models thus constructed for complexity and creativity as shown in Table \ref{quality_student_models}.

\begin{table}[ht!]
\caption{The quality of students’ models.}
\label{quality_student_models}
\begin{tabular}{p{0.05\linewidth}  p{0.135\linewidth}  p{0.135\linewidth} p{0.135\linewidth}p{0.05\linewidth}  p{0.135\linewidth}  p{0.135\linewidth} p{0.135\linewidth}}  
    \hline
   & \# of Hypo & Complexity & Creativity & & \# of Hypo & Complexity & Creativity \\
    \hline
     P1 & 3 & 9 & 2 & P9 & 2 & 7 & 5  \\ \hline
     P2 & 4 & 9 & 3 & P10 & 4 & 12 & 5  \\\hline
     P3 & 3 & 7 & 5 & P11 & 2 & 5 & 5  \\\hline
     P4 & 2 & 3 & 2 & P12 & 3 & 5 & 5  \\\hline
     P5 & 4 & 7 & 6 & P13 & 2 & 7 & 5  \\\hline
     P6 & 2 & 3 & 4 & P14 & 3 & 8 & 5  \\\hline
     P7 & 2 & 7 & 2 & P15 & 2 & 3 & 3  \\\hline
     P8 & 2 & 7 & 5 &  &  &  &   \\\hline
 
    \hline\\
\end{tabular}
\end{table}

The complexity of a model was calculated by adding the number of components in the conceptual model and the total number of relationships among the components. For example, Table \ref{quality_student_models} shows a high complexity model by Participant 10 and a low complexity model by Participant 4. The high complexity model (left) has 6 components and 6 links (complexity=12). The low complexity model(right) has 2 components and 1 link (complexity=3). If a student constructed multiple models, then the model with the highest complexity score was selected for that student.

The creativity score is concerned with whether the models contained “novel” components and relationships. The creativity score was calculated by adding the number of valid and novel components and relationships in the model. For example, if Model A and Model B each has three components, the model that has distinctive components–1 “Predator,” 1 “Prey,” and 1 “Social facto”–is considered to be more creative than the model that has redundant components–2 “Predators,” and 1 “Prey.” Thus, in Table 5, the creativity score of P14 is much higher than P1 (P14=5; P1=2) despite of the same number of hypothesis.

\subsection{Discussion}
\subsubsection{The Number of Hypotheses and the Complexity }
\begin{figure}[ht!]
    \centering
    \begin{minipage}{\textwidth}
        \centering
        \includegraphics[width=\textwidth]{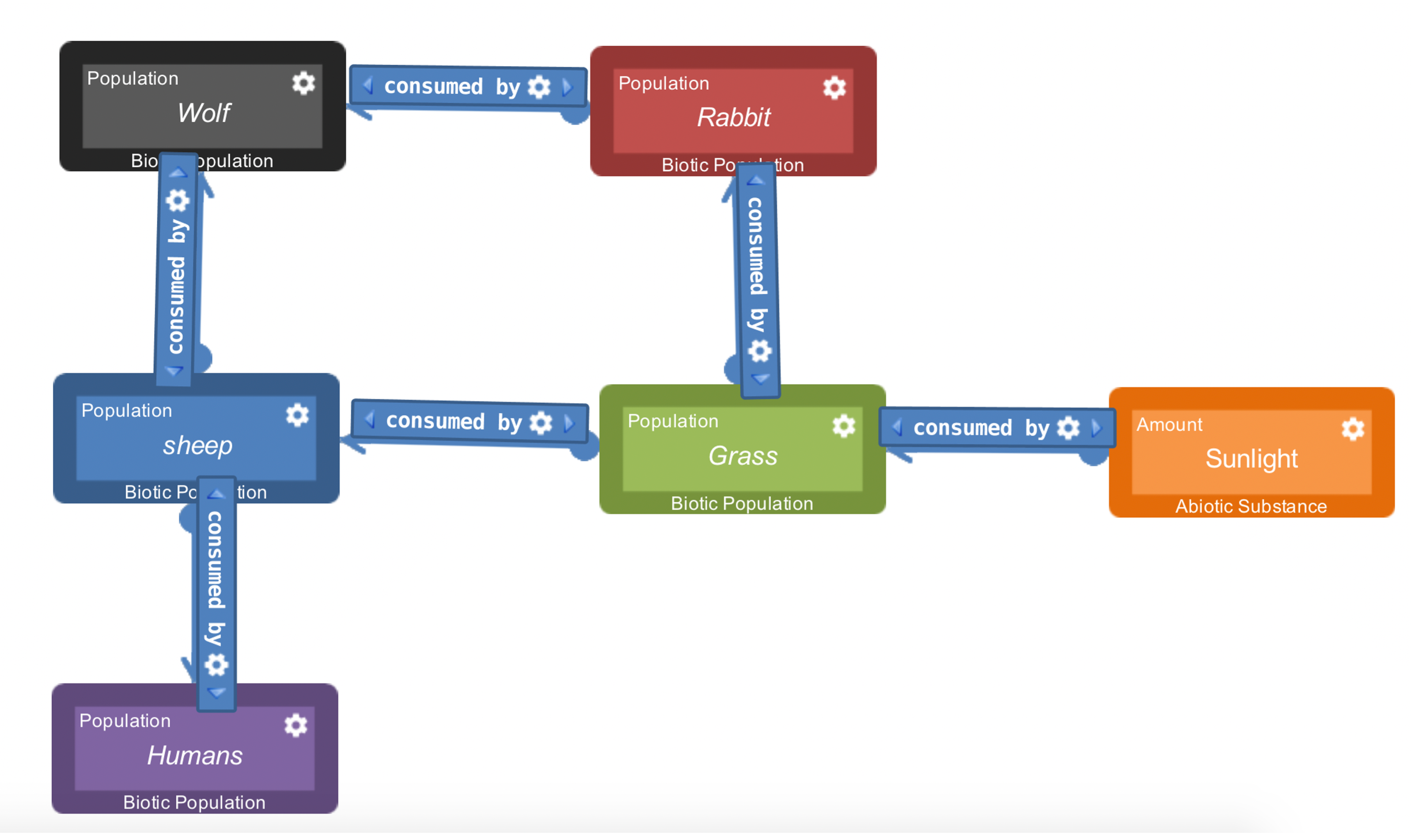}
        {\footnotesize (1) High Complexity Model (P10).\par}
    \end{minipage}\hfill
    \begin{minipage}{\textwidth}
        \centering
        \includegraphics[width=0.7\textwidth]{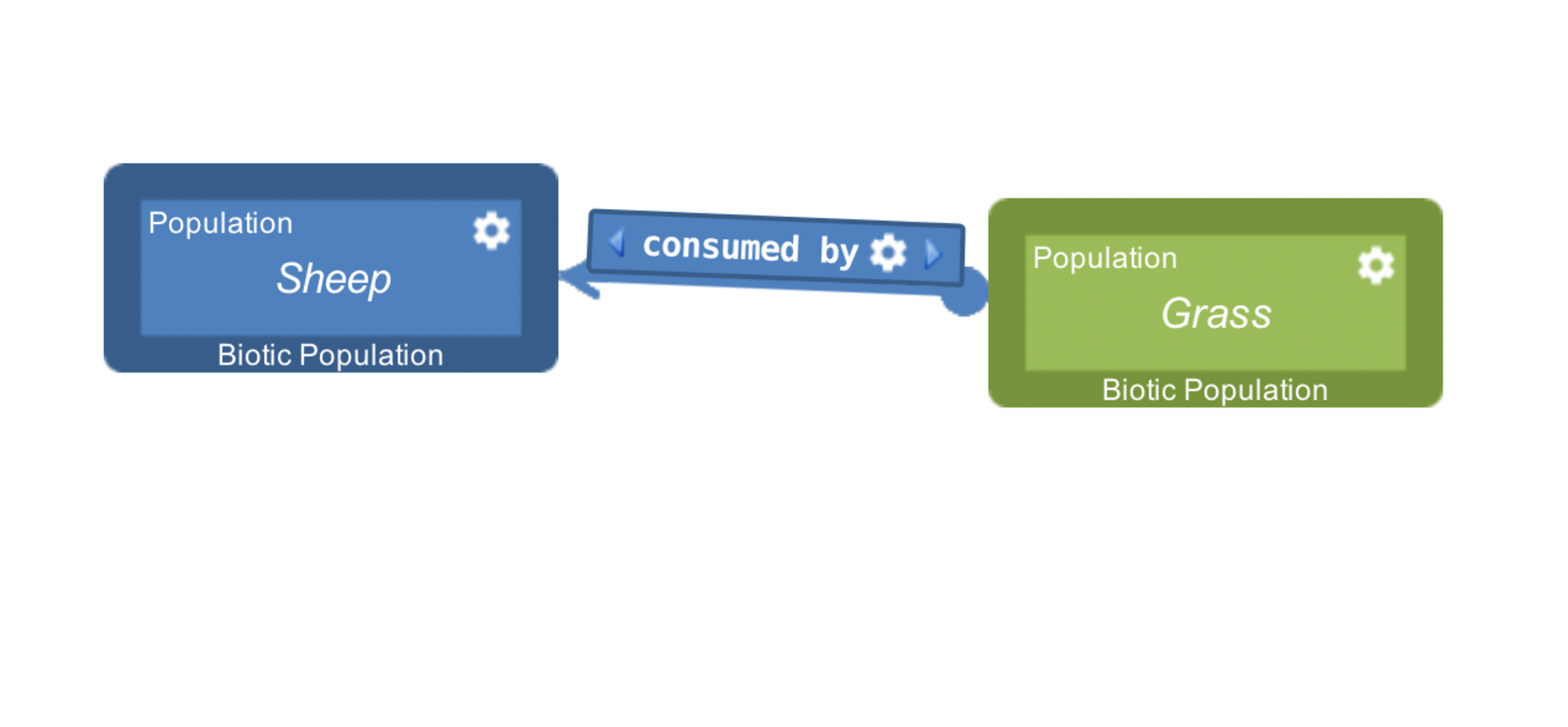}
        \\
        {\footnotesize (2) Low Complexity Model (P4).\par}
    \end{minipage}
    \caption{The example models of high and low complexity scores.}
    \label{high_low_complex_models}
\end{figure}

The number of hypotheses generated and the complexity of models constructed has a strong correlation (r=0.66). This means that the participants who generated many hypotheses are also likely to build complex models. For example, as shown in Figure \ref{high_low_complex_models}, the more complex model (left), has several hypotheses for the decline in sheep population: it could be an increase in the wolf population; an increase in the rabbit populations; an increase in human population; a decrease in the grass; or a decrease in sunlight. On the contrary, the less complex model (right) manifests only one hypothesis: a decrease in the grass.

\subsubsection{Complex Models and Creative Models}
Complex models are not always the most creative (r=0.18). While complex models have many components and relationships, complexity does not necessarily relate to the creativity of models. Some participants created a complex model with the same type of components. For example, P1 only used predators (Bear and Wolf) to construct three hypotheses: 1) Bear 2) Wolf 3) Bear and Wolf, which is considered to be less creative according to our criteria (see Table 5).

\subsubsection{Models and Learnings}
Gain in learning biological knowledge is related to the complexity of constructed models (r=0.29) whereas understanding the process of modeling is related to creative models (r=0.43). This means that the students who gained more than average in learning biological knowledge were better than average at building complex models; similarly, the students who improved in learning about the modeling process were better at making creative models.

\subsubsection{Models and EOL}
Use of EOL helped students build complex models. The students who answered that they used the EOL frequently were found to come up with multiple hypotheses and build complex models (r=0.38; r=0.26). We conjecture that information about the relationships between predator, prey, and competitors in the EOL knowledgebase may have led to the construction of more complex models. Further, EOL may have provided some students with new hypotheses for model construction. One participant mentioned in the post-test survey: “the page for sheep mentions diseases, which is why I made a model for parasites and infection.”


\begin{table}[ht!]
\scriptsize
\caption{The example hypotheses of high and low creativity scores.}
\label{student_hypotheses}
\begin{tabular}{p{0.47\linewidth}  p{0.47\linewidth}}  
    \hline
    \makecell{
    1) The sheep were eating grass, and the \\
    humans were eating sheep, and then there \\
    were wolves that were eating both the  \\
    humans and sheep.\\
    2) The second was that there was a \\
    disease and it was killing off the sheep.\\
    3) The third hypotheses was that there was \\
    a new plant that was killing the grass \\
    which was the sheep form of food and \\
    additionally, that there was a virus \\
    that was targeting the sheep as well.\\} & 
    \makecell{
    1) As the bear population increased, the \\
    sheep population would decrease because the \\
    bear preys on the sheep.\\
    2) As the wolf population increased, the \\
    sheep population would decrease the wolf \\
    preys on the sheep.\\
    3) As the bear and wolf populations increased,\\
    the sheep population would decrease because \\
    both the bear and wolf get their energy  \\
    directly from consuming the sheep.}\\
    \hline
    \makecell{
    Components: Predator, Prey, Disease/Virus \\
    Relationships: Consume, Parasite of} &
    \makecell{
    Component: Predator Relationship: Consume} \\
    \hline
    High Creativity Hypothesis (P14) & Low Creativity Hypothesis (P1) \\
    \hline

\end{tabular}
\end{table}

\section{Related Work}
Cognitive theories of scientific discovery postulate that science makes progress through inquiry-based model construction, evaluation, and revision \cite{clement2008creative}\cite{nersessian2010creating}. When scientists observe an abnormal or atypical phenomenon that cannot be explained with current knowledge, it forms a question in their minds. They construct models to develop hypotheses that can explain the phenomenon. As they evaluate their models, hypotheses not supported by data are rejected. They focus on the most plausible hypothesis and revises it to make it more accurate. Novice scientists such as student and citizen scientists too construct, revise, and update their knowledge through inquiry-based modeling \cite{sins2005difficult}\cite{schwarz2009developing}. However, scientific modeling, is cognitively challenging for novice scientists in part because of their limited domain knowledge \cite{hogan2001cognitive}, and and thus they requite cognitive assistance in model construction, evaluation, and revision \cite{sins2005difficult}\cite{schwarz2009developing}.

A variety of computational modeling tools have been developed to provide cognitive supports for novice scientists \cite{vanlehn2013model}. VERA provides a visual syntax for representing conceptual models of ecological phenomena, which includes concepts and relationships among concepts \cite{novak2008theory}\cite{white1990causal}. It uses the semantics of Component -Mechanism -Phenomenon (CMP) models that are composed of components of a system, causal interactions among the components, and phenomenon emerging from the interactions. The ontology of components and their interactions in VERA’s conceptual models is derived from Encyclopedia of Life (EOL). VERA uses agent-based simulations for evaluating the conceptual models. VERA borrows the ontology of components parameters for simulating the behaviors, such as life span, offspring count, reproductive maturity, etc., from Smithsonian's EOL. In addition, it presets the simulation parameters for the species available in EOL. This is especially important for novice modelers because they struggle with deciding the values to use due to limited knowledge and understanding about the ecological systems. 

Other than the MILA-S system, Co-Lab \cite{van2005co} and PROMETHEUS \cite{bridewell2006interactive} are the two systems closest to VERA. Co-Lab is a collaborative modeling environment where groups of early learners can conduct experiments via computer simulations to facilitate inquiry-based learning in natural sciences. Learners construct their models in Co-Lab by selecting pre-defined qualitative relations and mathematical formulas to express their hypotheses. On the other hand, PROMETHEUS is a modeling environment for more advanced modelers that enables them to visualize the structure of models, run them as simulations, and examine their predictions in earth and life sciences \cite{bridewell2006interactive}. PROMETHEUS uses qualitative differential equations for describing causal processes. For example, the connection between the variable, nasutum, and the process, nasutum\_decay indicates that the variable appears on the left-hand side of a differential equation within that process. Although PROMETHEUS helps visualize the structure of the models, the user has to construct the models by writing in its programming language.

VERA uses NetLogo for agent-based simulation \cite{wilensky1999thinking} because of its focus on ecology. The agent-based simulations enable VERA to specify the behaviors of various individuals that make up the system and give rise to emergent phenomena. Co-Lab and PROMETHEUS, on the other hand, use equation-based modeling that consists of a set of equations and executions to evaluate them. In the Co-Lab’s model editor, learners construct their models by selecting pre-defined qualitative relations and mathematical formulas to express their hypotheses. PROMETHEUS uses a process model language to specify variables, equations, and coefficients. VERA uses a visual language to specify conceptual models, without having to write in a programming language or express mathematical equations, and then automatically generates agent-based simulations.  

In the course of model revision, there are motivational and conceptual advantages to giving novice modelers actual data to improve their models \cite{hogan2001cognitive}. For instance, PROMETHEUS recommends possible changes that the users can make to improve their models to better fit actual data. Co-Lab also enables learners to compare their models to data from the simulation, which feeds their decision to revise the models. VERA does not currently support such cognitive assistance for model revision, which is left as future work. 

\section{Conclusion}
Inquiry-based modeling is central to scientific research. However, engagement in scientific modeling requires domain knowledge as well as quantitative skills. The Virtual Ecological Research Assistant (VERA) is an interactive learning environment that supports inquiry-based modeling for novice scientists such as student and citizen scientists. VERA builds on our previous work that provided a visual language for conceptual modeling and an AI compiler for automatic generation of agent-based simulations in support of model construction, evaluation, and revision. VERA leverages the Encyclopedia of Life to help users construct conceptual models and set the values of simulation parameters.

We conducted two studies with college-level biology students using VERA for modeling ecological phenomena. Both experiments demonstrate gains in ecological knowledge. We also found that the use of EOL helped students create more complex models. While the pre- and post-test comparisons indicate a gain in content knowledge, we did not observe similar gains on questions pertaining to conceptual and simulation modeling. We posit three main reasons. First, the learning might not actually have happened. Second, the questions might be not have been adequate in capturing students’ abilities or knowledge of modeling. Third, the pre- and post-test method might not be appropriate tool for capturing the learning that may have happened. The pre- and post-test comparison method may be useful in measuring content knowledge that can be learned in a short time, but not equally useful in measuring gains in modeling skills.

Several aspects of the present study invite additional research. For example, we found that prior student exposure to world-building games (such as SIMS and Minecraft) is correlated with gains in learning about ecology. We asked this question precisely because we surmised that background knowledge in simulation games may help students in using agent-based simulations and thus learning from them. In addition, we were not able to see how certain features of VERA affect the quality of students’ models and their experience in modeling. For example, how does the number of model revisions relate to students’ learning and modeling quality? We also need to compare participants’ performance with and without cognitive scaffolding in VERA to see if it actually helped students perform better in the post-tests and improve their models. Understanding the precise nature of this relationship requires additional research.

\subsubsection{Acknowledgements} At Georgia Tech, we thank David Joyner and Taylor Hartman for their contributions to MILA-S, and Abbinayaa Subrahmanian, Christopher Cassion and Pramodith Ballapuram for their contributions to VERA. At Smithsonian Institution, we thank Katja Schulz for her contributions to EOL. This research is supported by an US NSF grant \#1636848 (Big Data Spokes: Collaborative: Using Big Data for Environmental Sustainability: Big Data + AI Technology = Accessible, Usable, Useful Knowledge!) and the NSF South BigData Hub. This paper was presented as a poster at the 7th Annual Conference on Advances in Cognitive Systems (ACS), Cambridge, Aug 2019.





%
%
%
\bibliographystyle{splncs04}
\bibliography{mypaper}
%




\end{document}